\documentclass[sigconf]%,anonymous]
{acmart}
\AtBeginDocument{%
  }

\setcopyright{acmlicensed}
\copyrightyear{2026}
\acmYear{2026}
\acmDOI{XXXXXXX.XXXXXXX}
\acmISBN{978-1-4503-XXXX-X/2018/06}

\usepackage[most]{tcolorbox}

\usepackage{tikz}
\usetikzlibrary{positioning,arrows.meta,fit,calc}

\usepackage[table]{xcolor}
\usepackage{booktabs}
\usepackage{array}
\usepackage{tabularx}
\usepackage{longtable}

\usepackage{caption}
\usepackage{subcaption}
\usepackage{adjustbox}
\usepackage{booktabs}

\usepackage[ruled,linesnumbered]{algorithm2e}
\usepackage{soul}

\usepackage{url}
\usepackage{paralist}

\usepackage{comment}
\usepackage{xurl}
\usepackage{hyperref}

\usepackage{paralist}
\usepackage{comment} 
\usepackage{mdframed}
\usepackage{url}
\usepackage{xcolor, soul} 
\usepackage{paralist}
\usepackage[normalem]{ulem}
\usepackage[most]{tcolorbox}

\newboolean{showcomments}
\setboolean{showcomments}{true}
\ifthenelse{\boolean{showcomments}}
{\newcommand{\nb}[2]{
  \fcolorbox{black}{yellow}{\bfseries\sffamily\scriptsize#1}
  {\sf\small\textcolor{teal}{\textit{#2}}}
 }
 
}
{\newcommand{\nb}[2]{}
 
}

\usepackage[normalem]{ulem}
\newboolean{showdiffs}
\setboolean{showdiffs}{true}
\ifthenelse{\boolean{showdiffs}}
{
    
    \newcommand\del[1]{\textcolor{red}{\sout{#1}}}

}
{
    
    \newcommand\del[1]{}
}

\begin{document}

\title{%HSE-oriented development process of autonomous systems for digital humanity: Research Directions
%Ethics as runtime-requirements and Ethics-based Negotiation within and among SAS: Challenges and Research Directions}
%From Rules to Reasoning: Ethics at Runtime in \\ Self-Adaptive Systems \\ OPPURE \\ 
%From Rules to Runtime Reasoning:\\ %Negotiating 
The Runtime Dimension of Ethics in Self-Adaptive Systems} %: \\From Rules to Runtime Reasoning}

\author{Marco Autili}
\email{marco.autili@univaq.it}
\affiliation{%
  \institution{University of L'Aquila}
  \city{L'Aquila}
  \country{Italy}
}

\author{Gianluca Filippone}
\email{gianluca.filippone@gssi.it}
\affiliation{
\institution{Gran Sasso Science Institute}
\city{L'Aquila}
\country{Italy} 
}

\author{Mashal Afzal Memon}
\email{mashalafzal.memon@univaq.it}
\affiliation{%
  \institution{University of L'Aquila}
  \city{L'Aquila}
  \country{Italy}
}

\author{Patrizio Pelliccione}
\email{patrizio.pelliccione@gssi.it}
\affiliation{
\institution{Gran Sasso Science Institute}
\city{L'Aquila}
\country{Italy} 
}

\renewcommand{\shortauthors}{Autili et al.}

\begin{abstract}
% Self-adaptive systems increasingly operate in close interaction with humans, often sharing the same environments and making decisions with ethical implications at runtime. Existing approaches typically rely on fixed, rule-based notions of ethics, which is embedded into the systems. This overlooks the fact that ethical preferences vary across individuals and groups. In human–system interaction settings, such diversity is unavoidable: humans may actively interact with a system or be indirectly affected by its behavior, each bringing distinct ethical expectations.
% This paper advocates a shift from static ethical rules to runtime ethical reasoning in self-adaptive systems. We argue that ethical preferences should be treated as runtime requirements that evolve with context and stakeholders, and that satisfying them requires explicit ethics-based negotiation. By negotiating ethical trade-offs at runtime, self-adaptive systems can align their behavior with the preferences of multiple humans involved in or affected by their operation. We outline key challenges and research directions toward enabling ethically adaptive and socially aware systems.

Self-adaptive systems increasingly operate in close interaction with humans, often sharing the same physical or virtual environments and making decisions with ethical implications at runtime. Current approaches typically encode ethics as fixed, rule-based constraints or as a single chosen ethical theory embedded at design time. This overlooks a fundamental property of human-system interaction settings: ethical preferences vary across individuals and groups, evolve with context, and may conflict, while still needing to remain within a legally and regulatorily defined \emph{hard-ethics} envelope (e.g., safety and compliance constraints).
This paper advocates a shift from static ethical rules to \emph{runtime ethical reasoning} for self-adaptive systems, where ethical preferences are treated as runtime requirements that must be elicited, represented, and continuously revised as stakeholders and situations change. We argue that satisfying such requirements demands explicit \emph{ethics-based negotiation} to manage ethical trade-offs among multiple humans who interact with, are represented by, or are affected by a system. We identify key challenges, ethical uncertainty, conflicts among ethical values (including human, societal, and environmental drivers), and multi-dimensional/multi-party/multi-driver negotiation, and outline research directions and questions toward ethically self-adaptive systems.

\end{abstract}




\maketitle

\section{Introduction}\label{sec:intro}

Autonomous and increasingly AI-enabled software systems are becoming pervasive in everyday human activities, operating in close interaction with individuals, communities, and institutions. From recommendation and decision-support systems to cyber-physical systems and autonomous agents, such systems no longer merely optimize technical objectives but actively shape human behavior, social practices, and environmental outcomes~\cite{pelliccione2024insights,suri2023software,friedman2013value}. As a consequence, traditional notions of correctness, performance, and reliability are no longer sufficient: contemporary software systems must also be engineered to behave in ways that are ethically acceptable, socially legitimate, and environmentally responsible~\cite{de2024engineering}.

%From a software engineering perspective, this evolution challenges long-standing assumptions about requirements stability and system correctness.
From a software engineering perspective, this evolution challenges traditional assumptions about how requirements are specified, integrated, and validated and how system correctness is assessed.
In particular, ethical concerns such as fairness, privacy, accountability, transparency, and sustainability are increasingly recognized as first-class concerns throughout the software lifecycle, from requirements engineering to architecture, development, and runtime operation~\cite{Spiekermann2023,pelliccione2023architecting,autili2025engineering}.

Nevertheless, most existing approaches operationalize ethics by embedding rules, constraints, or ethical principles at design time, often in the form of fixed policies, priority schemes, or preselected ethical theories~\cite{bremner2019proactive,tolmeijer2020implementations,feng2024normative,dennis2016formal,townsend2022pluralistic,liao2023jiminy,winfield2014towards,Winfield2019,alidoosti2022incorporating}. %\pat{add references, for instance from machineethics}
These approaches implicitly assume that ethical expectations are sufficiently stable, homogeneous across individuals, and foreseeable to be resolved before deployment. %\gian{I added ``across individuals'' to highlight that we are referring to individual's preferences, not ethical dimensions in general like the ones listed above}
In practice, however, ethical expectations are inherently pluralistic and dynamic. They vary across individuals, stakeholder groups, cultures, and contexts, and they evolve over time as social norms, regulations, and situational factors change~\cite{10.1145/3635715}. 

Moreover, self-adaptive systems frequently operate in multi-stakeholder environments, where some humans directly interact with the system, while others are indirectly affected by its behavior. In such settings, ethical expectations may diverge or conflict, making it unrealistic to assume that a single, fixed ethical configuration can remain appropriate throughout the system lifecycle.
A useful conceptual lens for understanding this tension is Floridi’s distinction between \emph{hard} and \emph{soft} ethics~\cite{floridi2018soft}. Hard ethics refers to non-negotiable constraints imposed by law, regulation, and safety requirements, defining an envelope of admissible system behavior. Soft ethics concerns the legitimate but underdetermined space of ethical choices that remains within this envelope, such as how to balance privacy against personalization, efficiency against environmental impact, or individual autonomy against collective benefit. While hard ethics constrains what systems may do, soft ethics governs how systems ought to behave in context-dependent and stakeholder-sensitive ways. Importantly, soft-ethical decisions cannot be prescribed by regulations alone, since no single behavior can be identified as universally ethically acceptable; rather, multiple ethically defensible behaviors may exist, depending on contextual factors and stakeholder perspectives that cannot be exhaustively encoded in advance.
% regulation alone, since they persist precisely because multiple ethically defensible options remain available.

\begin{tcolorbox}[
  colback=white,
  colframe=blue!50,
  boxrule=0.1pt,
  left=6pt,right=6pt,top=4pt,bottom=4pt,
  sharp corners,
  fonttitle=\bfseries,
  title=Running example
]
\emph{Throughout this paper, we use an autonomous environmental monitoring drone as a running example. The drone adapts its flight paths, sensing modalities, and data collection strategies at runtime to balance monitoring accuracy, battery consumption, disturbance to wildlife, privacy concerns of nearby communities, and regulatory constraints. \\
While aviation safety rules and airspace regulations constitute hard ethical constraints, trade-offs between data quality, privacy, and ecological impact fall into the soft-ethics space and may legitimately vary across stakeholders and situations.}
\end{tcolorbox}

Building on this example, we argue that ethical preferences within the soft-ethics space should be treated as \emph{runtime requirements}. Rather than being fixed once-for-all at design time, such requirements must be monitored, interpreted, negotiated, and revised as part of the system’s adaptation loop. This view aligns naturally with self-adaptive systems engineering, which already provides mechanisms for monitoring, analysis, planning, and execution (e.g., MAPE-K loops). %However, ethical requirements differ from traditional quality attributes as they are \ins{subject to uncertainty (hereafter referred to as \textit{ethical uncertainty})} \chg{often uncertain, value-laden,}{, in that they are not fully known, individual- and context-dependent,} potentially conflicting, and may not admit crisp satisfaction criteria or total ordering.
However, ethical preferences can not be simply reduced to traditional requirements, as they are influenced by human and societal values or judgments instead of being purely technical, objective, or admitting a stable and universally agreed-upon satisfaction criteria. They are individual- and context-dependent, potentially conflicting, and may not admit crisp satisfaction criteria (hereafter we refer to this as \textit{ethical uncertainty}).
% \pat{Be careful with this sentence. Also other quality attributes are subject to uncertainty. I don't like much also the strange connection made by the sentence between requirements and quality attributes. I would focus only on the last part, the subjectivity, as follows.  ``However, we cannot consider ethics as traditional quality attributes as they are influenced by human and societal values or judgments than being purely technical, factual, or objective, they are individual- and context-dependent,
% potentially conflicting, and may not admit crisp satisfaction criteria (hereafter
% referred to as ethical uncertainty)." }
% \gian{I agree. I changed it a bit by saying that ethical preferences are not simply requirements. In the previous sentence we say that they can not be fixed once for all at design time, and that's reason why we need to adapt them. Now we further extend the problem: they are influenced by subjectivity and they do not admit a single interpretation. So this means that they are different, and this is also the main difference between ethical uncertainty and classical epistemic uncertainty} \marco{Ok, I also agree with both. We can implement this change, good :-)}
% \gian{I made a tiny change now, to connect with the negotiation (now we say ``do not admit an universally agreed-upon satisfaction criteria -> negotiation is needed at runtime)}

We contend that explicit \emph{ethics-based negotiation at runtime} is a key missing capability for self-adaptive systems operating in ethically pluralistic environments. Negotiation provides structured mechanisms to surface preferences, manage conflicts, and reach agreements among multiple parties. Yet, existing automated negotiation techniques predominantly focus on economic utilities or resource allocation and assume well-defined, commensurable preferences~\cite{kiruthika2020lifecycle,khemakhem2020agent,memon2025systematic}. Extending these techniques to ethical reasoning raises novel conceptual, representational, and assurance challenges, particularly when negotiation outcomes must remain within hard-ethics constraints and be explainable to affected stakeholders.

This paper advances the thesis that self-adaptive systems should move \emph{from rules to runtime reasoning} by explicitly negotiating ethical trade-offs within a hard-ethics envelope. We do not advocate replacing regulation or design-time ethical analysis; rather, we argue that these mechanisms must be complemented by runtime reasoning capabilities that account for evolving contexts and stakeholder diversity. Our contributions are threefold: (i) we frame ethical preferences as runtime requirements for self-adaptive systems, (ii) we analyze the resulting engineering and socio-technical challenges, and (iii) we outline research directions and research questions that define a research agenda for ethically adaptive and socially aware systems.

\paragraph{Paper organization.}
The remainder of this paper is organized as follows. Section~\ref{sec:literature} reviews related work on ethics in software and autonomous systems, value-sensitive design, runtime requirements, and automated negotiation, highlighting the limitations of design-time approaches. Section~\ref{sec:challenges} identifies the core challenges of ethics at runtime, focusing on ethical uncertainty, conflicts among human, societal, and environmental values, and the need for ethics-based multi-$\ast$ negotiation, illustrated through a running example. Section~\ref{sec:research_directions} derives research directions and research questions that operationalize these challenges into a coherent research agenda. Section~\ref{sec:challenges_to_rqs} explicitly maps challenges to research questions, clarifying their structural relationships. Finally, Section~\ref{sec:conclusions} concludes the paper and outlines directions for future work.

\section{State of the art}
\label{sec:literature}

%\marco{I have added sections below... following text to be aligned} 
This section presents the state of the art on ethical values, ethical considerations in autonomous systems, and negotiation approaches for achieving trade-offs.

\subsection{Ethical values} 
%\mashal{This is done now to me..}
%\marco{Check if aligned as I said and also extract challenges ... check also commented text} 
Values are fundamental principles that guide human behavior, influencing decision-making processes, interactions, and societal structures~\cite{DBLP:conf/sigsoft/MougoueiPHSW18,shahin2022operationalizing,bardi2009structure}. They serve as the underlying drivers of ethical considerations and cultural norms. In the context of human-societal and environmental considerations, values shape both individual and collective actions, influencing how systems and technologies are designed and implemented to act on behalf of humans~\cite{han2022aligning}.

The notion of values has been articulated in different ways. 
Certain studies focus on the introduction of human ethical values in the context of software architecture design~\cite{alidoosti2022incorporating,shahin2022operationalizing,DBLP:conf/sigsoft/MougoueiPHSW18,mougouei2020engineering,perera2020study,perera2019towards,nurwidyantoro2023integrating}, while other studies focus on various types of values~\cite{alidoosti2023stakeholder,ferrario2016values,liscio2022values,rokeach1973nature,maio2016psychology}. Besides software engineering, other fields such as artificial intelligence and human computer interaction have also emphasized the importance of incorporating values into system design~\cite{han2022aligning,chatila2019ieee,van2020embedding,trusilo2021ethical,floridi2018soft,floridi2019establishing,henschke2024pluralism}.

Consequently, several frameworks for extracting values have been proposed, including Schwartz’s Theory of Basic Human Values, the Value Sensitive Design (VSD) framework, etc.~\cite{jurkiewicz2004values,graham2013moral,schwartz2012overview,friedman2013value,feather1995values,rokeach1973nature,maio2016psychology,icse_BennaceurHNZ23}. These frameworks attempt to formalize values by capturing how they manifest in human behavior, governance, and technological systems. %Studies in various fields, such as software design~\cite{bersani2023architecting,winter2019advancing,whittle2019your,mougouei2020engineering}, artificial intelligence~\cite{han2022aligning,chatila2019ieee,van2020embedding}, and philosophy~\cite{trusilo2021ethical,floridi2018soft,floridi2019establishing,henschke2024pluralism}, have emphasized to incorporate values into autonomous systems. 
However, due to the subjective and uncertain nature of ethical values, it remains challenging to precisely define, operationalize, and integrate values into autonomous systems as a one-size-fit solution~\cite{le2009values,pommeranz2012elicitation,saket2021putting,DBLP:conf/sigsoft/MougoueiPHSW18,alidoosti2022incorporating,han2022aligning}.

%operationalizing values that are derived at runtime in a structured manner remains challenging~\cite{le2009values,pommeranz2012elicitation,saket2021putting,DBLP:conf/sigsoft/MougoueiPHSW18}.

%This is due to the subjective and uncertain nature of values, as each individual has different priorities. 
Moreover, due to their varying priorities, individuals may have values that potentially conflict with those of others. In practical scenarios, a multitude of values are relevant to each decision-making situation, and individuals assign different levels of importance to each of these values (i.e., each of us holds different value preferences)~\cite{liscio2022cross,siebert2022estimating}.
%Thus, integrating the dynamic and potentially conflicting values within systems remains a significant challenge.
%
%
%A key challenge in embedding values into systems is the inherent subjectivity and contextual dependence of values~\cite{alidoosti2022incorporating,han2022aligning}. 
%In practical scenarios, a multitude of values are relevant to each decision-making situation, and individuals assign different levels of importance to each of these values, (i.e., each of us holds different value preferences). Extracting human ethical values is a challenging task. 
%%Different disciplines perceive values through distinct lenses e.g., software engineering may prioritize fairness and transparency, while environmental studies may emphasize sustainability and ecological integrity. Despite efforts to integrate values into these domains, there is no universally accepted method for encoding, measuring, or enforcing values effectively.
%
%Another critical challenge is the presence of value conflicts as individuals and entities hold different perceptions of values~\cite{han2022aligning}. 
Thus, resolving these conflicts remains another challenge. Recent studies highlight the need for interdisciplinary approaches to address the challenges associated with value identification~\cite{inverardi2019ethics,inverardi2022ethical,Autili2019}. However, operationalizing uncertain ethical values derived at runtime and resolving conflicts among them remains unaddressed.

\subsection{Ethics in Autonomous Systems}\label{sec:ethics-in-autonom-sys}  
%\mashal{I will remove this section and add only ethical values, or maybe combine this with the that. }

%\marco{Mashal, as I told you last night, I have added the following. Check if to shorten and reuse text somewhere else if needed, e.g., for challenges} \mashal{I have reduced it.. }

Ethical considerations in autonomous systems have gained significant attention in recent years, as autonomous systems that make decisions on behalf of their users are likely to be more efficient than humans~\cite{suri2023software,jedlickova2024ensuring}. Consequently, the development of such systems has attracted the interest of the research community, leading to the birth of the field of ``\emph{Machine ethics}''~\cite{guarini2013introduction}, which is a combination of computational logic and moral philosophy~\cite{tolmeijer2020implementations}. Over the years, various studies have discussed fundamental principles of developing systems that account for ethics, including Moor's four levels of systems ranging up to those which are capable of human-like ethical
reasoning~\cite{picard2000affective,moor2006nature,wallach2008moral,gunkel2012machine,tolmeijer2020implementations}. In addition to these basic principles, various ethical theories including deontology~\cite{sep-ethics-deontological}, consequentialism~\cite{sep-consequentialism}, and others, have been introduced that systems can follow to guarantee their ethical conduct~\cite{huang2022overview}. However, the uncertainty of different ethical theories makes it difficult to identify a single ethical theory that should be taken into account while developing such systems~\cite{awad2018moral,bogosian2017implementation,nallur2019ethics}. %\mashal{Here, we have an idea to introduce research direction: to introduce models that consider ethics/HSE without needing to account for these theories, we already have it, just keeping this comment to be sure, we are in right direction.}

~\cite{autili5221280ethics}Building on this, the studies in~\cite{cardoso2021implementing, bremner2019proactive, winfield2014towards, Winfield2019} propose approaches for integrating an ethical theory as guiding principles within the planning mechanisms of autonomous systems. Before simulation, the planner evaluates each action based on the selected ethical framework, determining whether the actions conform to its principles. For example, actions that prioritize human safety (as in Asimov’s laws~\cite{asimov1941three}) are preferred and considered ethical. However, these approaches primarily emphasize rule adherence rather than accounting for dynamic ethical values stemming from human, societal, and environmental (HSE) drivers~\cite{autili5221280ethics}. This presents a need to engineer systems that must integrate evolving HSE values into their mechanisms, ensuring the system's alignment with dynamic ethical considerations at runtime.

Similarly, an approach for an artificial moral agent is presented in~\cite{liao2019building, liao2023jiminy}, where the authors propose a framework that incorporates the moral values of different stakeholders to guide ethical decision-making. In these studies, the systems use rules to derive a single ethical theory from different stakeholder values, which aligns with our approach. However, in these studies, the rules are predefined and given to the system to identify a single theory, whereas we focus on enabling the system to internally navigate trade-offs between dynamic values that evolve over-time and encompass not only individual but also societal and environmental considerations. The study in~\cite{townsend2022pluralistic} introduces approaches for translating normative principles into explicitly formulated practical rules to ensure that autonomous systems align social, legal, ethical, empathetic, and cultural (SLEEC) rules. The studies in~\cite{feng2024analyzing,sinem2025specification,feng2024normative,feng2023towards,TroquardSIPS24} further introduce methods to identify conflicts and semantic relationships between these rules. However, these studies consider principles and does not account for dynamic human, societal, and environmental values and ethical considerations that evolve at runtime. Moreover, several studies have emphasized the importance of developing systems that are ethically aware~\cite{alidoosti2022incorporating,moor2006nature,allen2006machine,tolmeijer2020implementations,huang2022overview,de2024engineering,inverardi2019ethics,bostrom2018ethics,floridi2019establishing,ryan2020artificial,alidoosti2023stakeholder}. However, approaches for integrating dynamic human, societal, and environmental values, enabling systems to handle their trade-offs at runtime, remains largely unexplored.

\subsection{Automated negotiation} 

Automated negotiation has become a key field of study, driven by the growing demand for systems capable of autonomously making decisions on behalf of individuals~\cite{luo2024survey}. Negotiation enables autonomous systems to communicate with each other and resolve conflicts by finding trade-offs to reach mutually acceptable agreements~\cite{baarslag2017automated,kiruthika2020lifecycle,bagga2022deep,zuckerman2013towards,baarslag2016learning,khemakhem2020agent}. Various approaches have been proposed throughout the years focusing on how agents learn about each other, exchange offers, and adopt strategic behaviors to optimize outcomes~\cite{memon2023automated,memon2025systematic}. %
The proposed approaches in existing studies for automated negotiation, mainly focus on economics, where agents negotiate prices, resource allocation, or financial gains~\cite{dimopoulos2019argumentation,rajavel2020agent,krohling2021context,kumar2022bilevel,bagga2022deep}. The primary goal of agents in these studies is to maximize their utility within constrained environments, such as e-commerce, supply chain management, and market-based interactions. While some studies address other aspects~\cite{baarslag2017automated,filipczuk2022automated}, such as exchanging privacy permissions to online services on behalf of their users, the exchange is done in terms of economic values where personal data is exchanged for monetary compensation.

A different perspective is presented in studies~\cite{baarslag2017value,le2018preference}, which focus on preference elicitation. These studies propose techniques where agents capture user preferences, enabling systems to consider them during the negotiation process. However, these approaches retrieve user preferences in terms of price and do not account for ethical preferences that could guide the system in making ethical decisions during negotiation.

Despite the existing research in automated negotiation, a crucial gap exists in proposing negotiation frameworks that consider integrating ethical values in the negotiation process to reach ethical agreements, beyond negotiating solely for price. As autonomous systems increasingly make decisions on behalf of humans and interact with other systems ~\cite{anderson2018artificial,waldman2019power}, it is important to develop negotiation mechanisms that incorporate ethical considerations in the decision-making process of autonomous systems~\cite{MemonSIA23}.

Developing systems that balance dynamic ethical values at runtime when making decisions necessitates a shift in the system engineering process~\cite{de2024engineering,autili2025engineering}. The engineering process must provide systems with components that can understand, represent, and reason about dynamic and potentially conflicting values, while also incorporating mechanisms that enable them to navigate value trade-offs when interacting with diverse stakeholders.

In addition, since dynamic ethical values are not considered part of negotiation, multi-dimensional negotiation of uncertain ethical values, where negotiation is conducted over multiple conflicting values, each prioritizing different objectives within a system, also remains unaddressed, as does the resolution of internal value conflicts to derive trade-offs that can subsequently be negotiated with other systems at runtime.

\subsection{From existing approaches to our perspective}

Existing research provides important foundations for embedding ethical considerations into autonomous and adaptive systems, but it rarely connects ethical reasoning with the core mechanisms of self-adaptation, such as runtime requirements management, conflict resolution, and decision assurance. In particular, current approaches offer limited support for negotiating ethical trade-offs at runtime across multiple stakeholders and dimensions, under conditions of uncertainty and regulatory constraints. This disconnect motivates a shift from static ethical rules to runtime ethical reasoning grounded in negotiation.

%\gian{None of the aforementioned papers deal with multi-level negoation and with negotiation among confling values during the runtime.}

%defining a computational representation of values, resolving conflicts between stakeholders, and ensuring alignment with ethical and societal principles. 
%As AI-driven systems take on greater roles in decision-making, enabling value-sensitive negotiation between systems will be essential to creating more responsible, transparent, and human-aligned technologies.
%\input{sections/example}
%\input{sections/new_challenges}
\section{Challenges for Ethics at Runtime} \label{sec:challenges}

Treating ethics as a runtime concern that is related to individual preferences fundamentally changes the nature, the reasons, and the scope of self-adaptation. Instead of adapting solely to technical conditions or performance goals, systems must adapt to evolving ethical expectations under uncertainty and conflicts. This shift raises challenges that cut across requirements engineering, runtime monitoring, decision-making, assurance, and human-system interaction. We structure these challenges into three interrelated dimensions. Throughout the section, we refer to the running example of an autonomous environmental monitoring drone that must balance monitoring accuracy, battery consumption, wildlife disturbance, privacy concerns of nearby communities, and regulatory constraints.

\subsection{Ethical uncertainty}
\label{subsec:ethical_uncertainty}

Ethical preferences are rarely fully known, precisely specified, or stable. 
They vary across individuals, are influenced by contextual factors and social interpretations, and may evolve throughout the system's lifetime, raising concerns that go beyond mere technical functionality~\cite{Spiekermann2023}.
This introduces multiple forms of uncertainty: uncertainty in eliciting stakeholder values, uncertainty in interpreting how abstract ethical principles apply to concrete situations, and uncertainty in predicting the long-term ethical consequences of actions.

\begin{tcolorbox}[
  colback=white,
  colframe=blue!50,
  boxrule=0.1pt,
  left=6pt,right=6pt,top=4pt,bottom=4pt,
  sharp corners,
  fonttitle=\bfseries,
  title=Running example
]
\emph{In the drone example, the acceptable level of wildlife disturbance or data collection granularity may depend on season, location, community sentiment, or evolving conservation priorities.}
\end{tcolorbox}

% any deviation from the unachievable ideal of completely deterministic knowledge of the relevant system

In this sense, ethics itself becomes a \emph{source of uncertainty}. We refer to this dimension as \emph{ethical uncertainty}, defined as the uncertainty arising from (i) the limited observability of stakeholders' ethical expectations at design time, (ii) their dependence on contextual and environmental conditions, and (iii) their plurality and potential ambiguity of interpretation. Unlike ``traditional'' forms of goal or requirement uncertainty~\cite{uncertaintyACSOS}, ethical uncertainty does not merely stem from incomplete or evolving specifications, but from the absence of a uniquely possible interpretation of ethical requirements, even when such requirements are explicitly stated.

From an engineering perspective, this implies that ethics cannot be treated as a conventional design-time requirement that can be fully elicited, specified, and validated during design-time requirements engineering, nor can it be exhaustively operationalized through fixed design and implementation choices.
Ethical uncertainty challenges traditional assumptions underlying runtime verification and adaptation.
First, ethical requirements may be fuzzy, probabilistic, or contested, making it difficult to define clear satisfaction conditions or adaptation triggers.
% Furthermore, systems must often act despite incomplete ethical information, raising questions about robustness, precaution, and justification /gian{non capisco che vuol dire :(}: this uncertainty can be only resolved at runtime, when the system interacts with humans (directly, as users and operators, or indirectly, as affected parties).
Furthermore, systems must often operate with incomplete ethical information, requiring runtime mechanisms to select among multiple ethically acceptable behaviors, reason on and revise ethical interpretations, and provide justifiable decisions.
Finally, it is unrealistic, if not impossible, to assume that ethical preferences remain acceptable for all users and contexts over the system’s lifetime, as they can not be completely and universally anticipated at design time.

This calls for a shift in what design-time requirements specify. If the system cannot be preconfigured with a single, fixed ethical configuration, addressing ethical uncertainty requires runtime models that can represent partial knowledge, reason under uncertainty, and support explanations of decisions taken in ethically ambiguous situations. This means that, at design time, instead of specifying requirements driving the system's ethics behavior, designers should instead specify \emph{ethical meta-requirements}, which prescribe \emph{how} the system acquires, represents, updates, and enforces ethical preferences at runtime (e.g., how feedback is incorporated, how conflicts are detected, how hard-ethics constraints delimit admissible decisions). In other words, design-time efforts must include explicit architectural provisions (``empty slots'') for runtime ethical reasoning and its operationalization.

\begin{tcolorbox}[
  colback=white,
  colframe=black!100,
  boxrule=0.6pt,
  left=6pt,right=6pt,top=4pt,bottom=4pt,
  sharp corners,
  fonttitle=\bfseries,
  title=Challenge CH1 (Ethical Uncertainty)
]
\emph{How can self-adaptive systems explicitly account for ethical uncertainty by shifting from design-time specification of ethics-driven behavior to runtime acquisition, management, and reasoning over ethical preferences (as runtime requirements)?}
\end{tcolorbox}

\subsection{Conflicting ethical values and HSE drivers}
\label{subsec:conflicts_hse}

Ethical conflicts arise both within and across stakeholders. A single stakeholder may value environmental protection, data accuracy, and operational efficiency, but only up to certain thresholds. Across stakeholders, conservation agencies, local residents, operators, and regulators may prioritize different and sometimes incompatible values. These conflicts, spanning across human, societal, and environmental (HSE) drivers, are often incommensurable.

\begin{tcolorbox}[
  colback=white,
  colframe=blue!50,
  boxrule=0.1pt,
  left=6pt,right=6pt,top=4pt,bottom=4pt,
  sharp corners,
  fonttitle=\bfseries,
  title=Running example
]
\emph{In the drone scenario, increasing monitoring frequency may improve environmental data quality while simultaneously increasing privacy intrusion or wildlife disturbance. Resolving such conflicts is not merely a matter of optimization; it involves normative judgments about acceptable trade-offs, often under time pressure and incomplete information. In addition, conflicts may be \emph{latent}, i.e., they can emerge only at runtime when contexts change (e.g., a new community event, seasonal sensitivity of wildlife, or changing social acceptance).}
\end{tcolorbox}

From a software engineering standpoint, the challenge is to detect emerging conflicts at runtime, explore feasible resolutions within hard-ethics constraints, and select outcomes that are not only technically feasible but also ethically defensible and socially legitimate. This requires moving beyond fixed, design-time prioritization rules, because (i) stakeholders may disagree on such predefined priorities, (ii) acceptable trade-offs may change with context, and (iii) multiple ethically defensible solutions may exist. A further implication is that conflict management must be auditable, which requires the system to be able to justify why a certain trade-off was selected, given the available ethical evidence and the applicable constraints.

\begin{tcolorbox}[
  colback=white,
  colframe=black!100,
  boxrule=0.6pt,
  left=6pt,right=6pt,top=4pt,bottom=4pt,
  sharp corners,
  fonttitle=\bfseries,
  title=Challenge CH2 (Conflicting Values and HSE Drivers)
]
\emph{How can self-adaptive systems detect and resolve runtime conflicts among non-commensurable ethical values (across human, societal, and environmental drivers), and justify the resulting trade-offs as ethically defensible and socially legitimate under hard-ethics constraints?}
\end{tcolorbox}

\subsection{Multi-$\ast$ negotiation}
\label{subsec:multistar_negotiation}

Runtime ethical adaptation requires negotiation mechanisms that go beyond classical formulations~\cite{memon2025systematic}. \emph{Multi-dimensional negotiation} is needed to handle heterogeneous ethical dimensions, such as privacy, fairness, and sustainability, without collapsing them into a single utility function. \emph{Multi-party negotiation} is required to account for the preferences of multiple stakeholders, including those indirectly affected by system behavior. \emph{Multi-driver negotiation} is required to account for potentially conflicting human, societal, and environmental drivers within the single system's behavior. %reflects the fact that ethical drivers must coexist with traditional adaptation drivers such as performance, cost, and safety.}{\emph{Multi-driver negotiation} reflects that a single system's behavior can be simultaneously shaped by multiple, potentially conflicting drivers such as human, society, and environment.} 

\begin{tcolorbox}[
  colback=white,
  colframe=blue!50,
  boxrule=0.1pt,
  left=6pt,right=6pt,top=4pt,bottom=4pt,
  sharp corners,
  fonttitle=\bfseries,
  title=Running example
]
\emph{In the running example, negotiation may be required to balance data quality against privacy and ecological impact, potentially involving operators, affected communities, and regulatory constraints. Importantly, negotiation is not only an outcome-oriented mechanism, since the \emph{negotiation process} itself is ethically relevant as well. Decisions about who participates, how stakeholders are represented, what information is disclosed, and which negotiation protocol is adopted affect legitimacy, fairness, and trust.}
\end{tcolorbox}

Integrating such negotiation mechanisms into self-adaptive systems raises additional engineering challenges, including scalability, convergence guarantees, stability of adaptation behavior, and runtime overhead. It also raises socio-technical challenges around participation and consent. For instance, how to account for affected stakeholders who cannot or do not actively participate, and how to ensure that negotiated outcomes remain within the hard-ethics envelope.

\begin{tcolorbox}[
  colback=white,
  colframe=black!100,
  boxrule=0.6pt,
  left=6pt,right=6pt,top=4pt,bottom=4pt,
  sharp corners,
  fonttitle=\bfseries,
  title=Challenge CH3 (Multi-$\ast$ Negotiation)
]
\emph{How can ethics-based multi-dimensional, multi-party, and multi-driver negotiation be integrated into self-adaptive systems so that negotiated outcomes converge, respect hard-ethics constraints, and are accountable to affected stakeholders?}
\end{tcolorbox}

\noindent
These challenges motivate the research directions in Section~\ref{sec:research_directions}, which focus on treating ethical preferences as runtime requirements, enabling reasoning under ethical uncertainty, managing conflicts across HSE drivers, supporting ethics-based multi-$\ast$ negotiation, and ensuring accountability and auditability.

\section{Research Directions and Research Questions}
\label{sec:research_directions}

Addressing the challenges outlined above requires a research agenda that bridges self-adaptive systems, requirements engineering, automated negotiation, and socio-technical ethics. In this section, we articulate a set of research directions (RDs) that operationalize the shift from static ethical rules to runtime ethical reasoning. Each direction is accompanied by research questions (RQs) that identify concrete, actionable problems and are illustrated using the running example of the autonomous environmental monitoring drone.

%\medskip
\subsection{RD1: Ethical Preferences as Runtime Requirements}
%\begin{tcolorbox}[
%  title=RD1: Ethical Preferences as Runtime Requirements,
%  colback=gray!5,
%  colframe=gray!60,
%  fonttitle=\bfseries
%]
A foundational step toward ethics at runtime is to model ethical preferences as \emph{first-class runtime requirements}. Unlike traditional functional or quality requirements, ethical preferences are inherently normative, context-dependent, and stakeholder-specific. Treating them as runtime requirements implies that they must be explicitly represented, monitored, revised, and reasoned about during system operation.
%\end{tcolorbox}

%\begin{tcolorbox}[
%  title=RD1 in the example,
%  colback=white,
%  colframe=gray!90,
%  fonttitle=\bfseries
%]

\begin{tcolorbox}[
  colback=white,
  colframe=blue!50,
  boxrule=0.1pt,
  left=6pt,right=6pt,top=4pt,bottom=4pt,
  sharp corners,
  fonttitle=\bfseries,
  title=Running example
]
\emph{In the drone example, acceptable levels of privacy intrusion or wildlife disturbance cannot be fixed once and for all. Privacy expectations may vary by location (e.g., near residential areas versus protected reserves), while acceptable disturbance thresholds may vary seasonally or with conservation priorities. Capturing such variability requires requirements models that go beyond static specifications and support uncertainty, contextualization, and plurality of stakeholders.}
\end{tcolorbox}

%\end{tcolorbox}

%\medskip

\subsubsection*{Research Questions (ethical preferences as runtime requirements)}

\begin{itemize}
  \item \textbf{RQ1.1} How can ethical preferences be specified as runtime requirements in such a way to preserve their normative meaning while remaining computationally tractable for adaptation and decision-making?
  \item \textbf{RQ1.2} How can runtime monitoring and inference mechanisms update ethical requirements based on contextual signals or stakeholder feedback without enabling manipulation or violating privacy?
\end{itemize}

%\medskip
\subsection{RD2: Reasoning and Assurance under Ethical Uncertainty}
%\begin{tcolorbox}[
%  title=RD2: Reasoning and Assurance under Ethical Uncertainty,
%  colback=gray!5,
%  colframe=gray!60,
%  fonttitle=\bfseries
%]
Ethical reasoning at runtime must explicitly account for uncertainty while still providing assurances about system behavior. Uncertainty arises at multiple levels: in eliciting ethical preferences, in interpreting how abstract values apply to concrete situations, and in assessing the consequences of adaptation decisions over time.

%\end{tcolorbox}
\begin{tcolorbox}[
  colback=white,
  colframe=blue!50,
  boxrule=0.1pt,
  left=6pt,right=6pt,top=4pt,bottom=4pt,
  sharp corners,
  fonttitle=\bfseries,
  title=Running example
]
\emph{In the drone scenario, uncertainty may concern the actual sensitivity of wildlife to disturbances, the expectations of nearby communities, or the long-term ecological impact of monitoring strategies. From an engineering perspective, this raises the challenge of making ethically informed decisions under partial or probabilistic information, while still ensuring predictable, explainable, and auditable behavior.}
\end{tcolorbox}

This research direction focuses on developing models and mechanisms that allow self-adaptive systems to reason under ethical uncertainty and to justify their decisions to human stakeholders, regulators, and auditors.

%\medskip
\subsubsection*{Research Questions (reasoning and assurance under uncertainty)}
%\textbf{Research Questions}

\begin{itemize}
  \item \textbf{RQ2.1} Which uncertainty representations (e.g., probabilistic, fuzzy, evidential) are most appropriate for ethical reasoning in self-adaptive systems, and how do they affect adaptation robustness and stability?
  \item \textbf{RQ2.2} How can systems generate meaningful explanations and justifications for ethically sensitive decisions made under uncertainty, particularly when trade-offs are unavoidable?
\end{itemize}

%\medskip
\subsection{RD3: Conflict Detection and Resolution across HSE Drivers}
%\begin{tcolorbox}[
%  title=RD3: Conflict Detection and Resolution across HSE Drivers,
%  colback=gray!5,
%  colframe=gray!60,
%  fonttitle=\bfseries
%]
Ethical conflicts are unavoidable in multi-stakeholder and multi-driver settings. Such conflicts may arise within a single stakeholder (e.g., valuing both privacy and data utility), across stakeholders (e.g., conservation agencies versus local communities), or across human, societal, and environmental (HSE) drivers.
%\end{tcolorbox}

\begin{tcolorbox}[
  colback=white,
  colframe=blue!50,
  boxrule=0.1pt,
  left=6pt,right=6pt,top=4pt,bottom=4pt,
  sharp corners,
  fonttitle=\bfseries,
  title=Running example
]
\emph{In the running example, increasing monitoring frequency may improve environmental data quality while simultaneously increasing privacy intrusion or wildlife disturbance. Resolving these conflicts is not merely a technical optimization problem, but a normative challenge that requires explicit reasoning about acceptable trade-offs.}
\end{tcolorbox}

% This research direction addresses the need for runtime mechanisms that can detect emerging ethical conflicts, characterize their nature, and support dynamic conflict-resolution strategies that are fair, transparent, and defensible.
% \mashal{What about replacing it with:} \ins{This research direction addresses the need for runtime mechanisms that can detect and model emerging ethical conflicts, and support the dynamic generation of conflict-resolution strategies that are transparent so their compliance with laws and hard ethical constraints can be verified, and fair, so that they do not unduly favor one stakeholder’s values over another’s in the absence of predefined rules or normative thresholds.}\pat{this section is tooooo long and confusing. Split it in two or three sentences. For instance: ``This research direction addresses the need for runtime mechanisms capable of modeling and detecting emerging ethical conflicts. Such mechanisms should support the dynamic generation of conflict-resolution strategies. These strategies must be transparent so that their compliance with legal requirements and hard ethical constraints can be verified. They must also be fair, ensuring that no stakeholder’s values are unduly favored over others when predefined rules or normative thresholds are absent.''}

This research direction addresses the need for runtime mechanisms capable of modeling and detecting emerging ethical conflicts. Such mechanisms should support the dynamic generation of conflict-resolution strategies. These strategies must be transparent so that their adherence to HSE values and their compliance with hard ethical constraints can be verified. They must also be fair, ensuring that no stakeholder’s values are unduly favored over others when predefined rules or normative thresholds are absent.

%\medskip
%\textbf
\subsubsection*{Research Questions (conflict detection and resolution across HSE drivers)}

\begin{itemize}
  \item \textbf{RQ3.1} How can self-adaptive systems detect and model emerging ethical conflicts within and across stakeholders and/or across human, societal, and environmental drivers at runtime?
  
  %\item \textbf{RQ3.2} What conflict-resolution strategies are appropriate for non-commensurable ethical values, and how can their \chg{fairness and legitimacy be assessed?}{fairness and transparency be assessed?}\pat{not sure here the focus is on transparency. When I read it I understood more: ``What conflict-resolution strategies are appropriate for non-commensurable ethical values, and how can their fairness and stakeholder acceptance be evaluated?''}
  \item \textbf{RQ3.2} What conflict-resolution strategies are appropriate for non-commensurable ethical values, and how can their fairness and stakeholder acceptance be evaluated?
\end{itemize}

%\medskip
\subsection{RD4: Ethics-Based Multi-$\ast$ Negotiation}
%\begin{tcolorbox}[
%  title=RD4: Ethics-Based Multi-$\ast$ Negotiation,
%  colback=gray!5,
%  colframe=gray!60,
%  fonttitle=\bfseries
%]
Negotiation provides a mechanism for managing ethical trade-offs at runtime. However, existing automated negotiation techniques mainly focus on price~\cite{luo2024survey,khemakhem2020agent,memon2025systematic}. Hence, they must be significantly extended or adapted to support ethical reasoning. Ethics-based negotiation is inherently multi-dimensional (involving heterogeneous values), multi-party (involving multiple stakeholders), and multi-driver (involving human, societal, and environmental aspects).

\begin{tcolorbox}[
  colback=white,
  colframe=blue!50,
  boxrule=0.1pt,
  left=6pt,right=6pt,top=4pt,bottom=4pt,
  sharp corners,
  fonttitle=\bfseries,
  title=Running example
]
\emph{In the drone example, negotiation may be required to balance monitoring accuracy against privacy and ecological impact, potentially involving operators, affected communities, and regulatory constraints. Moreover, the negotiation process itself raises ethical questions about participation, representation, and information disclosure.}
\end{tcolorbox}

%\end{tcolorbox}

This research direction focuses on integrating ethics-based negotiation into self-adaptive systems in a way that supports convergence, scalability, and explainability.

\begin{table*}[htbp]
\centering
\footnotesize
\begin{tabular}{p{0.15\linewidth} p{0.35\linewidth} p{0.42\linewidth}}
\toprule
\textbf{Challenge} 
& \textbf{Research Questions} 
& \textbf{Rationale} \\
\midrule
\textbf{CH1: Ethical Uncertainty} 
& RQ1.1, RQ1.2 (ethical preferences as runtime requirements); \newline \newline
  RQ2.1, RQ2.2 (reasoning and assurance under uncertainty). 
& Ethical uncertainty stems from the impossibility of fully specifying ethical preferences at design time. The associated RQs address how ethical preferences can be represented, monitored, and revised as runtime requirements, and how systems can reason about and justify adaptation decisions under incomplete, evolving, and uncertain ethical information. \\
\midrule
\textbf{CH2: Conflicting Ethical Values and HSE Drivers} 
& RQ3.1, RQ3.2 (conflict detection and resolution across HSE drivers); \newline \newline
  RQ4.1, RQ4.2 (multi-dimensional ethical negotiation).
& Conflicts among human, societal, and environmental values require mechanisms to detect and resolve non-commensurable trade-offs at runtime. The mapped RQs focus on identifying emerging ethical conflicts and on enabling multi-dimensional reasoning and negotiation to balance competing values within hard-ethics constraints.\\
\midrule
\textbf{CH3: Multi-$\ast$ Negotiation} 
& RQ4.1, RQ4.2 (multi-dimensional ethical negotiation); \newline \newline
  RQ5.1, RQ5.2 (accountability and auditability of negotiated outcomes). 
& Multi-$\ast$ negotiation introduces complexity across dimensions, stakeholders, and adaptation drivers. The corresponding RQs investigate how ethical negotiation can scale to multi-party and multi-driver settings, while ensuring that negotiated outcomes remain explainable, auditable, and accountable over time. \\
\bottomrule
\end{tabular}
\caption{Mapping between Challenges (CH) and Research Questions (RQs)}
\label{tab:ch_to_rq}
\end{table*}

%\medskip

%\textbf
\subsubsection*{Research Questions (multi-dimensional ethical negotiation)}

\begin{itemize}
  \item \textbf{RQ4.1} How can automated negotiation techniques be adapted to reason over heterogeneous ethical dimensions while ensuring convergence and interpretability?
  \item \textbf{RQ4.2} How can negotiation protocols ensure legitimate stakeholder participation, including indirectly affected parties, without making runtime adaptation infeasible?
\end{itemize}

%\medskip
\subsection{RD5: Accountability and Auditability of Runtime Ethics}
%\begin{tcolorbox}[
%  title=RD5: Accountability and Auditability of Runtime Ethics,
%  colback=gray!5,
%  colframe=gray!60,
%  fonttitle=\bfseries
%]
Systems reasoning about and enforcing ethical requirements at runtime must be accountable and auditable. This is particularly critical in regulated or safety-critical domains, where organizations must demonstrate compliance with legal and ethical requirements.

%\end{tcolorbox}

\begin{tcolorbox}[
  colback=white,
  colframe=blue!50,
  boxrule=0.1pt,
  left=6pt,right=6pt,top=4pt,bottom=4pt,
  sharp corners,
  fonttitle=\bfseries,
  title=Running example
]
\emph{In the drone scenario, accountability and auditability require that runtime ethical adaptations, such as altering flight paths to reduce wildlife disturbance or modifying sensing to address privacy concerns, are accompanied by explicit records of the ethical preferences, constraints, and contextual factors that motivated them. These records must demonstrate continuous compliance with hard regulatory constraints while making soft-ethical trade-offs transparent and inspectable over time. RD5 focuses on mechanisms that enable such traceability and post hoc justification, ensuring that evolving ethical behavior remains defensible to regulators and stakeholders.}
\end{tcolorbox}

\begin{figure*}[htbp]

\centering
%\makebox[.5\textwidth]{
\begin{tikzpicture}[
    node distance=1.0cm and 0.8cm,
    every node/.style={font=\scriptsize},
    box/.style={
        draw, rounded corners,
        align=center,
        minimum height=0.55cm,
        inner sep=2pt
    },
    % arrow/.style={->, thick}
    arrow/.style={-, thick}
]

\node (anchor) at (0,0) {};

% --------- Challenges (top row) ----------
\node[box, right=0.45cm of anchor, minimum width=2.6cm] (ch1)
{CH1\\Ethical Uncertainty};
\node[box, right=4.6cm of ch1, minimum width=3.2cm] (ch2) {CH2\\Conflicting Ethical Values (HSE)};
\node[box, right=1 of ch2, minimum width=2.8cm] (ch3) {CH3\\Multi-$\ast$ Negotiation};

% --------- Research Directions (middle row) ----------
\node[box, below=of anchor, minimum width=2.8cm] (rd1) {RD1\\Runtime Ethical\\Requirements};
\node[box, right=of rd1, minimum width=3.0cm] (rd2) {RD2\\Reasoning \& Assurance\\under Uncertainty};
\node[box, right=of rd2, minimum width=3.0cm] (rd3) {RD3\\Conflict Resolution\\(HSE)};
\node[box, right=of rd3, minimum width=3.2cm] (rd4) {RD4\\Ethics-based\\Multi-$\ast$ Negotiation};
\node[box, right=of rd4, minimum width=2.6cm] (rd5) {RD5\\Accountability \&\\Auditability};

% --------- Research Questions (bottom row) ----------
\node[box, below=of rd1, minimum width=2.8cm] (rq1) {RQ1.x\\Runtime Ethics};
\node[box, right=of rq1, minimum width=3.0cm] (rq2) {RQ2.x\\Ethical Uncertainty};
\node[box, right=of rq2, minimum width=3.0cm] (rq3) {RQ3.x\\Conflicting Values};
\node[box, right=of rq3, minimum width=3.2cm] (rq4) {RQ4.x\\Ethical Negotiation};
\node[box, right=of rq4, minimum width=2.6cm] (rq5) {RQ5.x\\Assurance};

% --------- Arrows CH -> RD ----------
\draw[arrow] (ch1.south) -- (rd1.north);
\draw[arrow] (ch1.south) -- (rd2.north);

\draw[arrow] (ch2.south) -- (rd3.north);
\draw[arrow] (ch2.south) -- (rd4.north);

\draw[arrow] (ch3.south) -- (rd4.north);
\draw[arrow] (ch3.south) -- (rd5.north);

% --------- Arrows RD -> RQ ----------
\draw[arrow] (rd1.south) -- (rq1.north);
\draw[arrow] (rd2.south) -- (rq2.north);
\draw[arrow] (rd3.south) -- (rq3.north);
\draw[arrow] (rd4.south) -- (rq4.north);
\draw[arrow] (rd5.south) -- (rq5.north);

% --------- Top framing ----------
%\node[above=0.4cm of ch2, font=\scriptsize\bfseries]
%{From Design-Time Ethical Rules to Runtime Ethical Reasoning};

% --------- Bottom framing ----------
%\node[below=0.4cm of rq3, font=\scriptsize, align=center]
%{Hard Ethics Envelope: Laws, Regulations, Safety Constraints};
\end{tikzpicture}%}

\caption{From Design-Time Ethical Rules to Runtime Ethical Reasoning}
%\caption{Compact overview of the relationship between challenges, research directions, and research questions}
\label{fig:compact_challenges_rds_rqs}
\end{figure*}
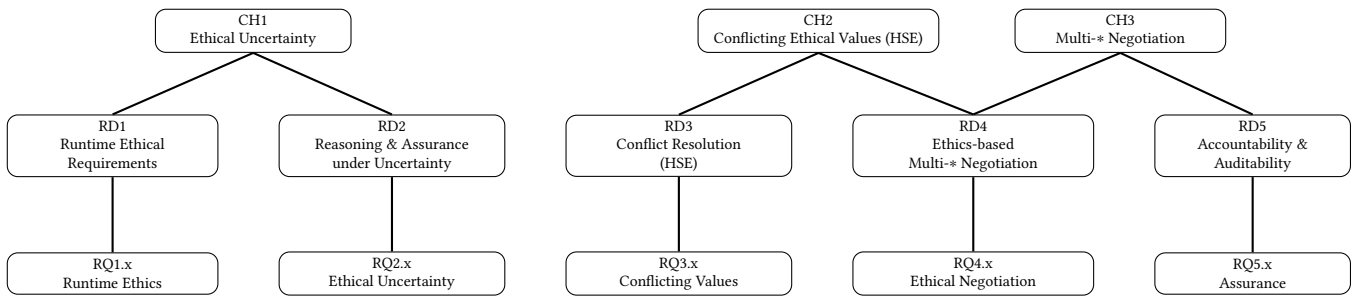

Due to the evolving, subjective, and runtime nature of ethical preferences, accountability becomes further complicated when the system needs to negotiate outcomes among conflicting values at runtime. %\gian{Then here, accountability is complicated because it is not just an adaptive setting, it is rather runtime/evolvable/subjective/context-bound}
This research direction addresses the need for evidence, traceability, and assurance cases that make ethical adaptation decisions inspectable and defensible over time.

% Due to the evolving, subjective, and runtime nature of ethical preferences, accountability becomes further complicated when the system needs to negotiate outcomes among conflicting values at runtime.

% As ethical preferences and decisions may evolve and be negotiated during system operation, runtime evidence, traceability, and assurance mechanisms are needed to ensure that ethical decisions remain inspectable and defensible over time.

% At runtime, accountability is further complicated when ethical preferences and decisions evolve or are negotiated during system operation, motivating the need for runtime evidence, traceability, and assurance mechanisms that make ethical decisions inspectable and defensible over time.

%\medskip

%\textbf
\subsubsection*{Research Questions (accountability and auditability of negotiated outcomes)}

\begin{itemize}
  \item \textbf{RQ5.1} What forms of runtime evidence and assurance cases are required to audit ethical adaptation decisions and negotiated outcomes?
  \item \textbf{RQ5.2} How can traceability from stakeholder inputs and contextual data to adaptation decisions be maintained over time?
\end{itemize}

\section{Mapping Challenges to Research Questions}
\label{sec:challenges_to_rqs}

The challenges identified in Section~\ref{sec:challenges} and the research directions and questions articulated in Section~\ref{sec:research_directions} jointly define the core contribution of this paper. To make this connection explicit and systematic, Table~\ref{tab:ch_to_rq} maps each challenge (CH) to the research questions (RQs) that are required to address it. Table~\ref{tab:ch_to_rq} should be read top-down. The resulting mapping clarifies how the proposed research agenda operationalizes the shift from static ethical rules to runtime ethical reasoning in self-adaptive systems.

Rather than presenting challenges and research questions as independent lists, the mapping highlights their structural relationship. Each challenge gives rise to multiple research questions, and several research questions jointly contribute to addressing more than one challenge. This reflects the intrinsically interdependent nature of ethics at runtime, where uncertainty, value conflicts, negotiation, and accountability cannot be treated in isolation.

% \begin{table}[t]
% \centering
% \footnotesize
% \caption{Mapping between Challenges (CH) and Research Questions (RQs)}
% \label{tab:ch_to_rq}
% \begin{tabular}{p{0.18\linewidth} p{0.72\linewidth}}
% \hline
% \textbf{Challenge} & \textbf{Addressed by Research Questions} \\
% \hline
% \textbf{CH1: Ethical Uncertainty} 
% & RQ1.1, RQ1.2 (ethical preferences as runtime requirements); \newline
%   RQ2.1, RQ2.2 (reasoning and assurance under uncertainty) \\
% \hline
% \textbf{CH2: Conflicting Ethical Values and HSE Drivers} 
% & RQ3.1, RQ3.2 (conflict detection and resolution across HSE drivers); \newline
%   RQ4.1 (multi-dimensional ethical negotiation) \\
% \hline
% \textbf{CH3: Multi-$\ast$ Negotiation} 
% & RQ4.1, RQ4.2 (ethics-based multi-dimensional and multi-party negotiation); \newline
%   RQ5.1, RQ5.2 (accountability and auditability of negotiated outcomes) \\
% \hline
% \end{tabular}
% \end{table}

\textbf{CH1 (Ethical Uncertainty)} motivates research questions that focus on how ethical preferences can be represented, monitored, and revised at runtime (RQ1.x), and how systems can reason and provide assurances under incomplete and uncertain ethical information (RQ2.x). In the running drone example, this concerns the system’s ability to cope with evolving privacy expectations, context-dependent wildlife sensitivity, and uncertain long-term impacts while still making justified adaptation decisions. Aligned with Table~\ref{tab:ch_to_rq}, Figure~\ref{fig:compact_challenges_rds_rqs} gives a compact overview of the proposed research agenda. Challenges related to ethics at runtime-ethical uncertainty, conflicting human, societal, and environmental values, and multi-$\ast$ negotiation are systematically mapped to research directions and research questions. 

%\del{The figure emphasizes the transition from static, design-time ethical rules to runtime ethical reasoning in self-adaptive systems, while explicitly acknowledging the role of non-negotiable hard-ethics constraints such as laws, regulations, and safety requirements.} \gian{I don't see how the figure emphasizes this. There is no mention of this transition in the figure.}\marco{I agree: that was the initial intention... remove the text.}

\textbf{CH2 (Conflicting Ethical Values and HSE Drivers)} is addressed by research questions that focus on detecting and resolving conflicts among human, societal, and environmental values (RQ3.x), as well as by negotiation mechanisms that allow systems to explore ethically admissible trade-offs across multiple dimensions through negotiation (RQ4.1). In the drone scenario, this corresponds to balancing data quality, privacy, and ecological impact when no single, universally optimal solution exists.

\textbf{CH3 (Multi-$\ast$ Negotiation)} requires research questions that extend automated negotiation to ethical settings involving multiple dimensions, stakeholders, and drivers (RQ4.x), while also ensuring that negotiated outcomes remain accountable, auditable, and explainable over time (RQ5.x). This is particularly relevant when multiple autonomous systems interact, each having internally derived ethical trade-offs that must be aligned with those of other systems and affected stakeholders.

Overall, this mapping demonstrates that the proposed research questions are not ad hoc but are systematically derived from the challenges posed by ethics at runtime. Together, they define a coherent research agenda that connects requirements engineering, self-adaptation, automated negotiation, and socio-technical ethics, and that is grounded in realistic, multi-stakeholder scenarios such as the running example used throughout the paper.

\section{Conclusions and Future Work}
\label{sec:conclusions}

Self-adaptive systems increasingly operate in ethically sensitive, multi-stakeholder environments, where decisions made at runtime can have significant human, societal, and environmental consequences. In such settings, ethics cannot be reduced to static, design-time rules or fixed priority schemes. This paper has argued for a fundamental shift \emph{from rules to runtime reasoning}, positioning ethical preferences as runtime requirements that are inherently uncertain, context-dependent, and potentially conflicting.

To support this shift, we have identified three core challenges for ethics at runtime: ethical uncertainty, conflicts among ethical values spanning across stakeholders and HSE drivers, and the need for ethics-based multi-$\ast$ negotiation. Building on these challenges, we articulated a set of research directions and research questions that collectively define a coherent research agenda. The explicit mapping between challenges and research questions highlights that addressing ethics at runtime requires coordinated advances across requirements engineering, self-adaptation, automated negotiation, and socio-technical assurance.

This paper does not propose a concrete solution, but rather establishes a structured foundation for future work in this space. As a next step, we plan to develop an extended body of work that systematically addresses the identified challenges by following the proposed research directions and empirically answering the associated research questions.

Methodologically, this future work will adopt a \emph{design-oriented and multi-method research approach}. First, we will follow a design science methodology to iteratively design, implement, and refine architectural and runtime mechanisms for ethical reasoning and negotiation in self-adaptive systems. This includes the definition of ethical meta-requirements, runtime models for ethical preferences, and negotiation mechanisms that operate under ethical uncertainty and hard-ethics constraints.

Second, we will ground and evaluate these mechanisms through empirical methods. Controlled experiments and simulation-based studies will be used to assess how systems behave under varying ethical preferences, stakeholder configurations, and contextual conditions, including the emergence of latent ethical conflicts. Case studies in representative domains, such as autonomous environmental monitoring systems, will be used to evaluate realism, scalability, and socio-technical validity. Where appropriate, user studies and stakeholder-in-the-loop evaluations will complement system-level experiments to assess perceived legitimacy, transparency, and trustworthiness of runtime ethical decisions.

Third, we will employ analytical and assurance-oriented methods to address accountability and auditability. This includes developing traceability mechanisms, runtime evidence collection, and assurance cases that demonstrate how negotiated ethical decisions remain within hard-ethics boundaries while adapting to soft-ethics variability. Together, these methods aim to provide not only functional correctness but also ethically grounded assurances for adaptive behavior.

%To reflect on how automated testing practices are reshaped by LLM-based systems we use McLuhan’s tetrad~\cite{mcluhan1988media}.
%McLuhan’s tetrad offers a holistic lens on technological change by examining what a technology enhances, obsolesces, retrieves, and reverses when pushed to its limits

By combining design, experimentation, and assurance in a systematic way, the planned extended work will operationalize the research agenda outlined in this paper. Ultimately, we envision this line of research contributing to the engineering of self-adaptive systems that are not only technically robust, but also ethically adaptive, socially legitimate, and accountable in the face of evolving human values and societal expectations.

% Autonomous systems are increasingly becoming an integral part of our daily lives, making it essential for their decision-making processes to incorporate ethical considerations. However, traditional software engineering focuses mainly on functional requirements and does not inherently address ethical concerns. To bridge this gap, this paper identifies several key challenges to integrate ethics into autonomous systems: (i) xxx \mashal{when we finalize them.}

% To address these challenges, we proposed research directions aimed at enabling autonomous systems to dynamically interpret, evaluate, and balance values alongside functional requirements at run-time. Moreover, when multiple autonomous systems interact, we emphasize the need xxxx during automated negotiation mechanisms that enable these systems to exchange offers, evaluate trade-offs, and reach agreements that reflect diverse ethical perspectives.

% % By advancing these research directions, we can lay the foundation for responsible, ethically aware autonomous systems that can operate effectively while respecting diverse stakeholder values
\section*{Acknowledgments}
\label{sec:acknowledgments}

This work has been partially funded by 
\begin{inparaenum}[(a)]
\item the MUR (Italy) Department of Excellence 2023 - 2027, 
\item the PRIN project P2022RSW5W - RoboChor: Robot Choreography, 
\item the PRIN project 2022JKA4SL - HALO: etHical-aware AdjustabLe autOnomous systems, 
\item the PRIN project 2022JAFATE - CAVIA: enabling the Cloud-to-Autonomous-Vehicles continuum for future Industrial Applications, and
\item Italian PNRR MUR Centro Nazionale HPC, Big Data e Quantum Computing, Spoke9 - Digital Society \& Smart Cities.
\end{inparaenum}

\bibliographystyle{ACM-Reference-Format}
\bibliography{ref}

\end{document}